\documentclass[twocolumn]{aa}
\usepackage{graphics,astron}
\title{ISO detection of a 60\,$\mu$m source near GRB\,970508}
\author{L.\,Hanlon\inst{1} \and R.J.\,Laureijs\inst{2} \and 
L.\,Metcalfe\inst{2} \and B.\,McBreen\inst{1}  \and
B.\,Altieri\inst{3} \and A.\,Castro-Tirado\inst{4,5} \and
A.\,Claret\inst{6}    
\and  E.\,Costa\inst{7} \and M.\,Delaney\inst{1,8}
\and M.\,Feroci\inst{7} \and F.\,Frontera\inst{9}  \and T.\,Galama\inst{10} 
\and J.\,Gorosabel\inst{4} \and P.\,Groot\inst{10} \and J.\,Heise\inst{11} 
\and M.\,Kessler\inst{2} \and C.\,Kouveliotou\inst{12} 
\and E.\,Palazzi\inst{9} \and J.\,van Paradijs$\dagger$\inst{10,13} 
\and L.\,Piro\inst{7} \and N.\,Smith\inst{14}
}
\institute{Department of Experimental Physics, University College Dublin,
  Belfield, Stillorgan Road, Dublin 4, Ireland 
\and ISO Data Centre, Astrophysics Division, ESA, Villafranca, Spain
\and XMM Data Centre, Astrophysics Division, ESA, Villafranca, Spain
\and Laboratorio de Astrof\'{\i}sica Espacial y F\'{\i}sica
Fundamental, Villafranca del Castillo, P.O. Box 50727, 
28080 Madrid, Spain
\and Instituto de Astrof\'{\i}sica de Andaluc\'{\i}a (IAA-CSIC), 
P.O. Box 03004, E-18080 Granada, Spain
\and Service d'Astrophysique, CEA/DSM/DAPNIA Saclay, Orme des
Merisiers, 91191 Gif-sur-Yvette C\'{e}dex, France
\and Istituto Astrofisica Spaziale, CNR, Roma I-00133, Italy
\and Stockholm Observatory, SE-133 36 Saltsj\"{o}baden, Sweden
\and ITESRE-CNR, Bologna, Italy
\and Astronomical Institut `Anton Pannekoek', University of
  Amsterdam, Amsterdam, The Netherlands
\and SRON Utrecht, The Netherlands
\and USRA at NASA/MSFC, Huntsville AL, USA
\and Physics Department, University of Alabama, Huntsville, USA
\and Department of Applied Physics and Instrumentation, 
Cork Institute of Technology, Cork, Ireland
}
\mail{lhanlon@bermuda.ucd.ie}
\date{Draft \today }
\thesaurus{03(13.07.1; 13.07.02; 13.09.1; 13.09.2)}
\titlerunning{ISO detection of a 60\,$\mu$m source near GRB\,970508}
\begin{document}
\maketitle
\begin{abstract}
The Infrared Space Observatory  observed the field of the
$\gamma$--ray burst GRB\,970508 with the CAM and PHT
instruments on May 21 and 24, 1997 and with PHT in three filters in November
1997. A source  at 60\,$\mu$m (flux in
May of $66\pm 10$\,mJy) was detected near the position of the host galaxy
of this $\gamma$--ray burst. The source was detected again in November
1997, at a marginally lower flux ($43\pm 13$\,mJy). A 
Galactic cirrus origin and a stellar origin 
for the emission can be ruled out on the basis of the infrared colours.
The marginal evidence for variability in the
60\,$\mu$m flux between May and November is not sufficient to warrant
interpretation of the source as transient fireball emission. However, 
 the infrared colours are physically reasonable if
attributed to conventional dust emission from a single blackbody
source. The probability of detecting a 60\,$\mu$m by chance in a PHT
beam down to a detection limit of 50\,mJy is $\sim 5\times 10^{-3}$.  
If the source is at the redshift of the host galaxy of the
$\gamma$--ray burst the fluxes and upper limits at 
wavelengths from 12\,$\mu$m to 170\,$\mu$m indicate it is an 
ultraluminous infrared galaxy (L$_{\rm ir} \sim 2\times
10^{12}$\,L$_{\sun}$). The star
formation rate is estimated to be several hundred solar masses per
year, depending significantly on model-dependent parameters. If this
source is associated with the host galaxy of GRB\,970508, progenitor 
models which associate GRBs with star-forming regions are favoured. 
\end{abstract}
\section{Introduction}
The recent discoveries of fading afterglows
to a number of $\gamma$--ray bursts (GRBs) have been precipitated by 
the accurate and prompt localisation capability of the two Wide Field Cameras
(WFC) aboard the `BeppoSAX'  
X-ray satellite (Boella et al. 1997; Piro et al. 1998a). 
\nocite{bb97,pcf+98}
The typical $\sim 3'$ radius error
circles, obtained in a matter of hours after the occurrence of the
GRBs, are easily covered by ground-based optical and radio
telescopes, allowing rapid and deep follow-up observations.  
The reduction of GRB error circles by BeppoSAX also made it
feasible to study the content of these regions at far-infrared
wavelengths with the European Space Agency's Infrared Space
Observatory\footnote{Based on observations with ISO, an ESA project with instruments
     funded by ESA Member States (especially the PI countries: France,
     Germany, the Netherlands and the United Kingdom) with the
     participation of ISAS and NASA.}, ISO \cite{ksa+96}. 
The error region of GRB\,970402 was the first 
to be rapidly surveyed at far-infrared wavelengths \cite{ajct98}. 
ISO observed this GRB error circle 55 hours, and
again 8 days, after the burst event and detected no new
transient sources down to a 5\,$\sigma$ limit of 0.14\,mJy
at 12\,$\mu$m and 350\,mJy at 170\,$\mu$m. Details of the ISO
observations of GRBs may be found in Delaney et al. (1999)
\nocite{dhm+99}.    
\section{Observations of GRB\,970508}
GRB 970508 triggered the GRB Monitor \cite{fro+97} of
BeppoSAX at 21:41:47 UT on May 8, 1997 \cite{cos+97}. 
The GRB localisation was subsequently improved by the BeppoSAX WFC2 detector  
to a $3'$ radius (99\% confidence) error circle \cite{hei+97}. 
Follow-up observations by the BeppoSAX narrow-field instruments (NFI) 
then detected a new, fading X--ray source, consistent with the GRB error 
circle, but  with an improved positional accuracy of 50$''$ radius
(Piro et al. 1998b).   
\nocite{phj+98} Within this error circle an optical transient (OT) was
discovered whose flux increased for the first two days and 
subsequently decayed \cite{bond97,dmk+97}. The decay phase was characterised 
by a power law dependence with time, $f(t) \propto t^{-1.1}$ days 
\cite{cgb+98}. 
 Near-infrared
(2.2$\mu$m) emission was also observed from the OT source between May
13.25 (K$_{\rm s}$ = 18.2$\pm 0.2$ mag) and May 20.21 (K$_{\rm s}$ =
19$\pm 0.3$ mag) \cite{cnm+98}. Spectra of the OT taken
 with the Keck II 10\,m telescope 
revealed the presence of [OII] emission and FeII and MgII   
absorption lines consistent with a source at a redshift of 0.835, 
providing the first direct  measurement of the distance to a GRB
(Metzger et al. 1997a, 1997b).  \nocite{mcc+97,mdk+97} 
Radio afterglow was also detected from a source
at the position of the OT \cite{fkn+97,bkg+98,gwb+98}. Observations
 by HST have shown the OT position to be remarkably 
well-centred with respect to the host galaxy \cite{fpg+99}. The host
 is late-type, with the characteristics of a blue compact galaxy
\cite{szb+98}. 

We report here on ISO observations of the field of GRB\,970508 in the weeks
following the GRB and on a follow-up observation made in November
1997. 
\section{ISO Observations}
The GRB\,970508 error circle was not initially visible with ISO  
due to orbital constraints but 
became  visible for seven short windows lasting from 3000 to 8500
seconds between May~\,20 and May~\,26, during which time 
target of opportunity observations were made with 
ISOCAM \cite{caa+96} and  ISOPHOT \cite{lka+96}. 
A second window was available from 
 October 25 to November 21 1997 and follow-up observations 
were then made with ISOPHOT. 
Details and results
from the CAM observations are presented elsewhere \cite{hmd+99}. 
The May observations were
centred at RA(2000): 6\,h\,53\,m\,46.7\,s, Dec(2000): 
+79$^{\circ}$\,16$'$\,02.0$''$, the centre of the X--ray afterglow
error circle \cite{pcf+98}. The 
November observations were centred on the position of the optical 
transient associated with the GRB \cite{bond97}. 

Raster maps of the NFI 50$''$ radius error circle were obtained 
with the PHT C100 and C200 detectors \cite{isophot94}.
The PHT C100 camera is a 3$\times$3 pixel Ge:\,Ga array, with a pixel
centre-to-centre distance of $46''$ in both directions. 
The C$\_60$ filter (reference wavelength at
60\,$\mu$m, width 23.9\,$\mu$m) and the C$\_90$ filter (reference wavelength at
90\,$\mu$m, width 51.4\,$\mu$m) are associated with
the C100 detector. Observations in these filters employed 
a 5$\times$5 position raster grid with  a $23''$ step size in both directions.
The grid was aligned with the orientation of the
detector array. The size of the area covered was therefore
$230''\times230''$. This mode of observing was selected to obtain a
fully sampled map with a high level of redundancy in the map centre
and to minimise the uncertainties due to cirrus confusion.
The redundancy was highest in the inner $46''\times46''$ where it was
25.  At 60 and 90\,$\mu$m the FWHM of the beam profiles (psf convolved
with pixel) are $41''$ and $47''$ respectively.

The PHT C200 camera is a 2$\times$2 pixel stressed Ge:\,Ga array, with a
pixel centre-to-centre
 separation of $92''$. A 5$\times$5 grid raster with a $46''$
step size in both directions was used for the C200 observations in the
C$\_160$ filter band (reference wavelength at 170\,$\mu$m, width 89.4\,$\mu$m). 
The area covered was $368''\times368''$ with a redundancy for the
inner $91''\times91''$ of 16.
\section{ISOPHOT Data Processing}
The ISOPHOT data were processed using the software package PIA version
7.3 \cite{gah+97}. All standard processing steps were included to obtain
the flux  per raster point per detector pixel before map
coaddition \cite{lkr+98}. This flux was taken to be 
the median value of all signals
measured during a given raster point integration thereby 
minimising any positive flux bias due to the relatively high number of
glitches, induced by ionising particles, in the data stream.  
The data were flatfielded by imposing the same integrated flux for all
detector pixel scans. 
This operation yielded a correction factor for each pixel without affecting
the total flux. The maps were constructed by coadding the flatfielded
fluxes per raster point per detector pixel in a regular image grid of
$23''{\times}23''$ image pixels which are aligned with the orientation of
the detector pixels.

Photometry on the  C$\_60$ and  C$\_90$ maps 
was carried out by integrating the flux in an area of
$5\times 5$ image pixels, minus the corner pixels, in the region of
interest and then subtracting the mean background 
flux derived from the remaining pixels in the map. The measured signal
was then scaled up to the full PHT beam-size. Since the maps from May
and November have different orientations
and map centres, the 60\,$\mu$m, 90\,$\mu$m and merged (combined
60\,$\mu$m and 90\,$\mu$m) maps were reprojected 
to have the position of the OT as map centre and with 
zero position angle. These maps were used to derive photometric fluxes.
\section{Results}
\begin{figure}[htbp]
\centering
\rotatebox{-270}{\resizebox{!}{9cm}{\includegraphics*{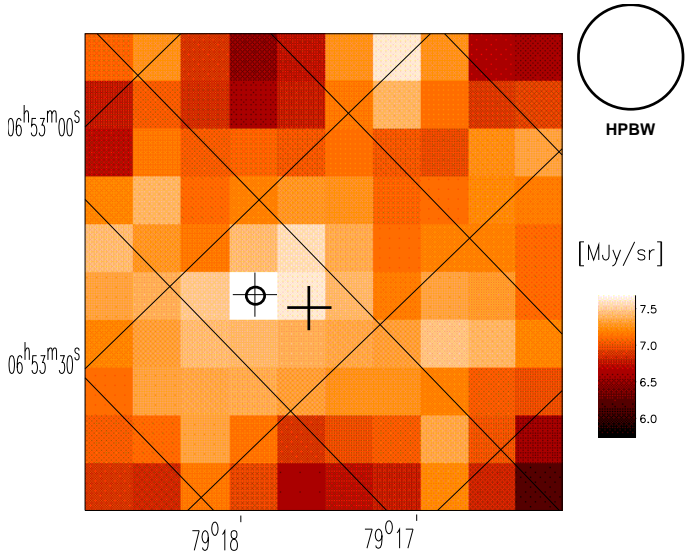}}}
          \caption{PHT C100 map using the
C\_60 filter, taken on May 24, 1997. The OT position (marked with a
cross) is consistent with the location of the C\_60 source (marked
with a circle) to within the HPBW of the C100 detector. The size of
the HPBW is shown on the side of the figure, along with the grey scale
values. }
          \label{pht_map555}
\end{figure}

\begin{figure}[htbp]
\centering
\rotatebox{-270}{\resizebox{!}{9cm}{\includegraphics*{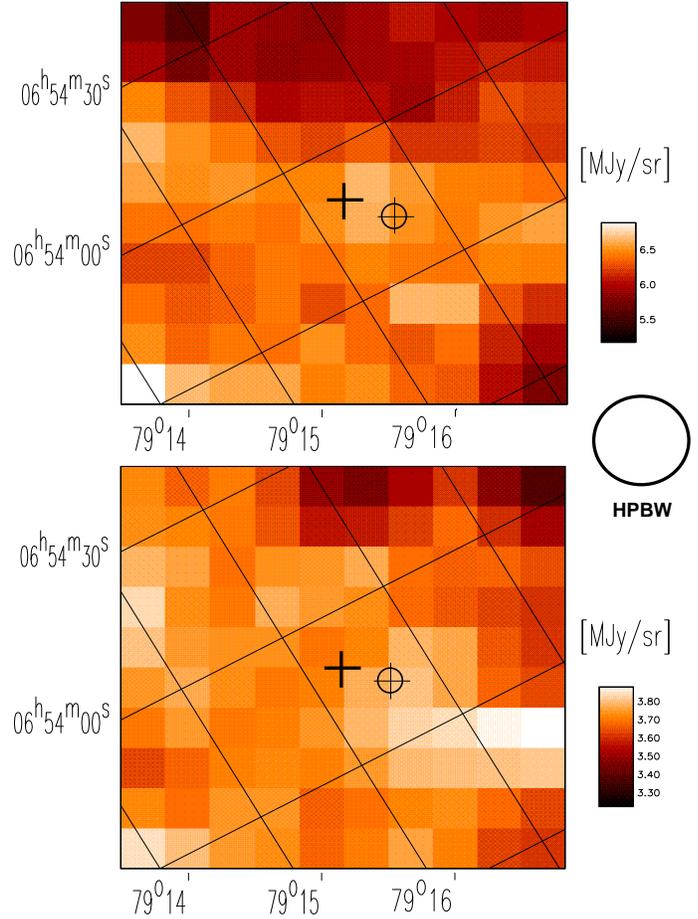}}}
          \caption{PHT C100 map using the
C\_60 filter (top) and the C\_90 filter (bottom), taken on November
11, 1997. The OT position is marked with a
cross while the position of the C\_60 source detected in May 1997 is marked
with a circle. A source is present in the C\_60 map which is consistent
with both the position of the OT and the C\_60 source detected in May
1997. Note: The orientation of these maps is different to that
in Fig.\,1, due to  differences in orientation and map centres between the May
and November observations.}
          \label{pht_map726}
\end{figure}

\begin{figure}[htbp]
\centering
\rotatebox{-270}{\resizebox{!}{9cm}{\includegraphics*{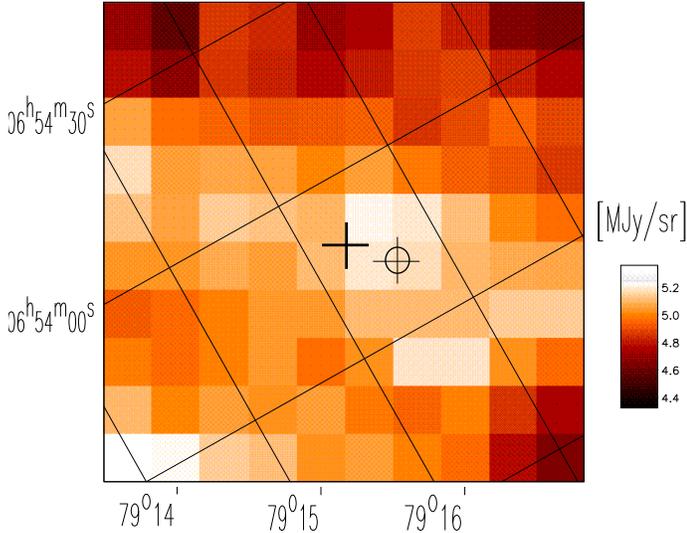}}}
          \caption{Merged 60\,$\mu$m and 90\,$\mu$m map from November
11, 1997. The OT position is marked with a cross. The 
local maximum near the position of the OT is marked with a circle. 
There are other sources near the edge of the map which are
brighter but have lower signal to noise due to the lower coverage in
those areas. }
          \label{pht_map_concat}
\end{figure}

A point source was detected at 60\,$\mu$m on May 24,  
consistent with the OT position, 
 given the uncertainty in the position of the peak
of the beam profile with respect to the geometric shape of a given
 pixel (Fig.~\ref{pht_map555}). For a relatively low
signal to noise detection such as this (see Table~\ref{isolog}), 
which is significant in only a few pixels
(not all 9), an offset of even a few arcsec can cause the mapping
routine to shift the source by an entire image pixel. However, it is
possible that the offset of the source relative to the OT position is
real and the consequences of this are discussed below.
Further observations were made with PHT in November 1997 to verify
the detection. 
The source was again detected at 60\,$\mu$m (albeit at a lower
significance level than in May) but no significant excess
at the position of the OT was detected at 90\,$\mu$m 
(see Fig.~\ref{pht_map726} and Table~\ref{isolog}). Although the
60\,$\mu$m flux is lower in November 
than in May, we do not regard this as providing
strong evidence for variability, since the two fluxes agree to within
2$\sigma$. The offset of the source relative to the OT is again
observed in the November map. 
In an attempt to improve the signal to noise ratio of the C100
observations from 
November 1997, the data from both filters were merged into a
single map (Fig.~\ref{pht_map_concat}). 
The 60\,$\mu$m and 90\,$\mu$m data were separately flatfielded and the
corrected fluxes per
pixel per detector and per filter were coadded on a common grid. The
fluxes per filter were weighted by their respective uncertainties.
The merged map has a noise level intermediate between the noise in the
individual maps and the significance of this is discussed below. 
In addition, the source is detected at a 
signal to noise ratio consistent with that of the 60\,$\mu$m map
alone. This suggests that the source is not significant at 
90\,$\mu$m. Upper limits ($3\sigma$), derived from the $170\,\mu$m 
observations in May and November, are given in Table~\ref{isolog}.
\begin{table}[htbp]
     \begin{center}
          \begin{tabular}{|c|c|l|l|}
\hline
UT Start - End   & Instrument & Filter & Flux Density  \\
                 &            &         &     (mJy) \\     
\hline
May 1997         &            &         &           \\
\hline
24.095 - 24.106  &  PHT  & C\_60 & $66\pm 10$  \\
24.107 - 24.114  &  PHT & C\_160 & $< 70$ \\
\hline
November 1997    &      &        &        \\
\hline
11.021 - 11.040  &  PHT  &C\_60  & $43\pm 13$ \\
11.041 - 11.065  &  PHT  & C\_90 & $19\pm 7$ \\
11.066 - 11.070  &  PHT  &C\_160  & $< 130$ \\
\hline
\multicolumn{3}{|c|}{Merged C$\_60$ plus C$\_90$} & \multicolumn{1}{l|}{$32\pm 9$} \\ 
\hline
          \end{tabular}
          \caption{Log of ISOPHOT observations of GRB\,970508. The PHT fluxes
            correspond to the reprojected maps (see text for details).}
          \label{isolog}
     \end{center}
\end{table}
\subsection{Assessment of uncertainties in the maps}
Per raster point, the integration times were 20\,s at 60\,$\mu$m in 
the May observation and 60\,s at 60\,$\mu$m and 90\,$\mu$m in the November
observations. Although the
integration time per raster point was three times longer at 60\,$\mu$m in the
second epoch, the noise levels per raster point in the two 60\,$\mu$m maps
are comparable (35\,mJy/beam in May, 31\,mJy/beam in November). 
The relative increase in the noise level of the  C\_60 measurement in 
November is most likely due to  
the fact that the observation was made at the end of a revolution
and was therefore subject to an increased glitch rate as the
satellite re-entered the radiation belts \cite{ksa+96}. 
The noise level in the C\_90
measurement was 17\,mJy/beam. Including the redundancy, the
predicted flux level in the map centre is about 12 mJy/beam at 60\,$\mu$m
and 6 mJy/beam at 90\,$\mu$m.  

The detection limit is also determined by the cirrus confusion noise 
($N$), which can be estimated from 
\begin{displaymath}
N({\rm mJy})=1.08 \left(\frac{\lambda}{100\mu{\rm
m}}\right)^{2.5} \times \left(\frac{I_{\nu}(\lambda)}{1 {\rm MJy/sr}}\right)^{1.5} \nonumber
\end{displaymath}

\noindent where $\lambda$ is the wavelength, 
$I_{\nu}$ is the background surface brightness level and a telescope
of diameter 60\,cm is assumed \cite{hb90}. 
The validity of the
formula is discussed in Herbstmeier et al. (1998) \nocite{hal+98}, 
and is probably better than an order of magnitude. The detection limit
 also depends on the absolute
calibration of the background surface brightness. The maps indicate
background surface brightness levels in the range 3-6\,MJy/sr which
correspond to cirrus confusion noise fluxes of 
N(60\,$\mu$m) $\sim$ 1.5--4.4\,mJy and N(90\,$\mu$m) $\sim$ 5.6--15.8\,mJy.
These numbers indicate that most of the noise measured at 90\,$\mu$m (7\,mJy
according to Table~\ref{isolog}) is probably due to cirrus structures.
However, this is not the case at 60\,$\mu$m. Therefore the noise
in the merged 60+90\,$\mu$m map is higher than the noise in the
90\,$\mu$m map, contrary to the predicted weighted uncertainty if the
noise is purely statistical. The presence of cirrus structures at 
90\,$\mu$m therefore imposes a
limit on the signal to noise which can be achieved. 

\section{Discussion}
\subsection{Origin of the 60\,$\mu$m emission}
Blink comparison of the November 60\,$\mu$m and 90\,$\mu$m maps
 (Fig.~\ref{pht_map726}) indicate a 
striking similarity in background structure between them.  A
correlation analysis confirms that there is a structure in the maps
(correlation coefficient 0.62) but that the random noise dominates
over the intrinsic variations in the background. The surface brightness
ratio (in MJy/sr)  determined from the maps is consistent
with that expected from cirrus in the Galaxy (I90/I60 = $2.03\pm
0.9$). It is
therefore plausible that the background structures in the maps are due
to cirrus. 
If the 90\,$\mu$m emission  at the position of the GRB is entirely
attributed to cirrus, then the maximum cirrus contribution to the flux
at 60\,$\mu$m is $\sim$10\,mJy. Subtracting this component from the
 observed fluxes in May and November, an averaged source flux
of 45$\pm 18$\,mJy is obtained. 
On the other hand, if the 60\,$\mu$m flux is attributed to
cirrus then the flux at 90\,$\mu$m should be $\sim100$\,mJy
which is clearly not the case. We can thus rule out a Galactic cirrus
origin for the source at 60\,$\mu$m.

A stellar origin for the 60\,$\mu$m source can also be excluded  on the basis
that the ISO colours are incompatible with stellar spectra over the
full range of spectral types \cite{skhl87,hs98}.

We can also exclude with reasonable confidence the possibility that the
 60\,$\mu$m source is entirely 
due to transient fireball emission from the GRB on the basis that (a) the
flux at 60\,$\mu$m on May 24 is about two orders of magnitude greater than the
extrapolated radio-optical afterglow spectrum on the same date and
(b) the source is detected again in November at a consistent flux
 level (to within 2$\sigma$). If
the 60\,$\mu$m emission followed the  t$^{-1.1}$ decay of
the afterglow, its flux should have decayed by a factor of roughly
15 by November. The relativistic shock scenario accounts quite
 well for the features of the 
broadband spectral energy distributions of GRB afterglows
and a new physical mechanism, beyond the scope of this paper,
 would have to be developed  
if the extraordinary excess at 60\,$\mu$m was to be  
connected to the fireball. We now argue instead that
the infrared colours we observe in the source are physically
 reasonable if attributed to conventional dust emission from a single
 blackbody source.

Modified blackbody spectra of the form 
\begin{math}
\nu^{n} \times {\rm B}_{\nu}{\rm (T)}, 
\end{math}
were fit to the range of 60\,$\mu$m/12\,$\mu$m colour ratios ($R$)
obtained from observations in May 1997 \cite{hmd+99}:
\begin{tabbing}
\rm  F60\,$\mu$m=66\,mJy/F12\,$\mu$m=0.065\,mJy \= $\Rightarrow$
$R \sim 1000$ \\
  F60\,$\mu$m=76\,mJy/F12\,$\mu$m=0.035\,mJy \> $\Rightarrow$ $R \sim 2200$ \\
  F60\,$\mu$m=56\,mJy/F12\,$\mu$m=0.095\,mJy \> $\Rightarrow$ $R \sim 600$ \\
\end{tabbing}
No colour corrections were incorporated in view of the errors on the flux
determinations. Estimated temperatures for modified blackbody spectra
 with $0 < n < 2$ were then obtained for these $R$ values
and predicted F90\,$\mu$m/F60\,$\mu$m ratios were calculated for
each derived dust temperature and assumed emissivity (Table~2). 
\begin{table}[htbp]
     \begin{center}
          \begin{tabular}{|c|c|c|c||c|c|c|}
\hline
$R$ & \multicolumn{3}{|c||}{T$_{\rm dust}$ (K)} &
          \multicolumn{3}{|c|}{Predicted F90\,$\mu$m/F60\,$\mu$m}  \\
\hline
    & $n=0$ & $n=1$ & $n=2$ & $n=0$ & $n=1$ & $n=2$ \\
\hline
600 &  91 & 81 & 70 & 0.79 & 0.58 & 0.44 \\
1000 & 87 & 76 & 68 & 0.83 & 0.61 & 0.45 \\
2200 & 82 & 72 & 65 & 0.86 & 0.64 & 0.47 \\
\hline
\end{tabular}
\caption{Modified blackbody spectral fits for different values of $R$,
yielding a range of estimated dust temperatures and predicted
F90\,$\mu$m/F60\,$\mu$m colours. 
The $R$ values were derived from the May 1997 CAM and PHT data.}
\end{center}
\end{table}
The temperatures for each dust model are well constrained. Moreover,
the predicted 90\,$\mu$m fluxes do not vary significantly for a given
$n$ even when $R$ changes by almost a factor of 4. These results also show
that subtracting a 10\,mJy cirrus
component from the 60\,$\mu$m flux does not significantly affect 
the conclusion.
Since no observations were made at 90\,$\mu$m in May 1997, due to orbital 
constraints, we cannot compare these predicted ratios with
observations at the same epoch.   
However, the fit results can be used to predict the 90\,$\mu$m flux in
November 1997 by appropriate scaling of the 60\,$\mu$m flux ratios from 
May (66$\pm 10$\,mJy) and November (43$\pm 13$\,mJy). Taking into
account the photometric errors, the ratio of the May to November
60\,$\mu$m flux is in the range 76/30 (2.5) to 56/56 (1). 
Multiplying the predicted F90\,$\mu$m/F60\,$\mu$m ratios 
in Table~2 by these scaling factors we
see that the predicted 90\,$\mu$m flux for November 1997 is consistent with
the observed 90\,$\mu$m value for all dust models when the scaling factor
is 2.5  and for the $n=2$ case when the scaling factor is 1.
Therefore it is not necessary to speculate that the infrared source
has varied between the two epochs because the ISO observations are consistent
with dust emission from a single blackbody with a temperature of $\sim
70$\,K and $n=2$.   
The predicted 170\,$\mu$m fluxes are always within the upper limits obtained
(Table~\ref{isolog}), being a factor of two lower than the 90\,$\mu$m
values.
 
\begin{figure}[htbp]
\centering
\resizebox{!}{7cm}{\includegraphics*{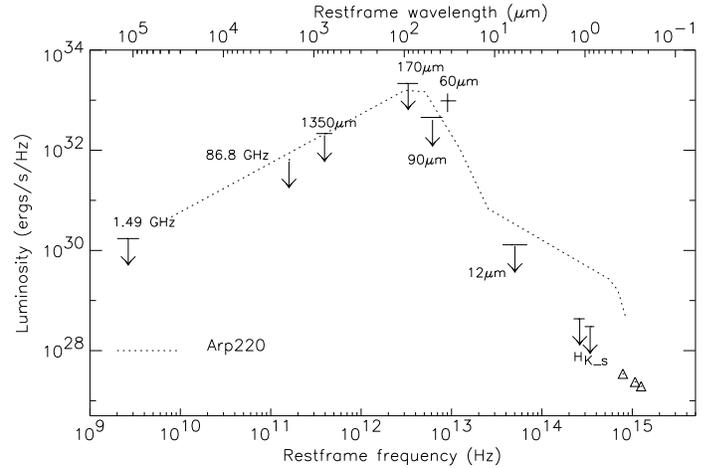}}
          \caption{Spectral energy distribution of the 60$\mu$m
source, assuming it is the host
            galaxy of GRB\,970508. Data and upper limits are from
Bloom et al. 1998; Pian et al. 1998; Smith et al. 1999; Galama et
al. 1998; Shepherd et al. 1998
{\protect\nocite{bdkf98,pfb+98,stp+99,gwb+98,sfkm98}} and the present work. 
The 12\,$\mu$m upper limit was obtained by
co-addition of the 2 CAM LW10 May observations. 
The C\_60 data point represents the
averaged flux from May and November. The SED of the ULIG Arp\,220 is
 shown for comparison purposes only.  }
          \label{sed}
\end{figure}

If the 60\,$\mu$m source is associated with the GRB host galaxy, at a
redshift of 0.835, limits can be placed on its bolometric luminosity. 
A broadband spectral energy distribution (SED) for the 60\,$\mu$m
source, assuming it is the host galaxy of the GRB,  
is shown in Fig.~\ref{sed}. The assumed cosmological parameters are
H$_0$=50\,km\,s$^{-1}$\,Mpc$^{-1}$ and q$_0$=0.5. 
If the upper limits across the spectral range
are included and the area under the SED is integrated  using a power
law interpolation between spectral points, 
an upper limit on the bolometric luminosity (0.4--1000\,$\mu$m) of  
$2.9 \times 10^{12}$L$_{\sun}$ is obtained. 
A value of $2.0 \times 10^{12}$L$_{\sun}$  is obtained if we consider 
only the ISO data points and upper limits as contributing towards the 
luminosity and clearly the emission is dominated by the infrared component.
Alternatively, the 60\,$\mu$m source may be associated with the
foreground absorbing system at z=0.767 in which case the infrared 
luminosity is $\sim 1.7 \times 10^{12}$L$_{\sun}$.
In either case the source is an
ultraluminous infrared galaxy (ULIG, L$_{\rm ir} > 10^{12}$L$_{\sun}$)
\cite{sm96}. Although ULIGs were discovered by the IRAS satellite in
our local universe (z $ <$ 0.2) \cite{ssn+86} a recent re-analysis of
the IRAS data has shown the existence of ULIGs out to z=0.529
\cite{wcsh98}. The extremely rare  hyperluminous infrared galaxies (HyLIG,
L$_{\rm ir} > 10^{13}$L$_{\sun}$) have been
detected to z=2.2 \cite{rbl+91,chl+94}. 

There are two galaxies within 5$''$ of GRB\,970508, either of which
may be responsible for the absorption system seen at z=0.767
\cite{pfb+98}. Due to the large PHT pixel size, it is possible that
one of these galaxies, rather than the GRB host, is the
counterpart of the 60\,$\mu$m source. The galaxy to the north-west of
the GRB, `G2' \cite{pfb+98} has the colours of a late-type spiral at
a redshift of 0.7, making it unlikely to be the counterpart, 
since IRAS detected no late-type spirals with 
${\rm  L}_{\rm ir} > 10^{11} {\rm L}_{\sun}$ \cite{sm96}. 
Galaxy `G1' to the north-east, has extremely blue colours 
indicative of a rapidly star-forming system \cite{pfb+98} and we
cannot rule out that it is the counterpart of the 60\,$\mu$m source. 

The prototypical ULIG is Arp\,220 with a far-infrared luminosity
$>10^{12}$L$_{\sun}$. The SED of Arp\,220 (Fig.~\ref{sed}) 
consists of a cold component, 
corresponding to a modified blackbody temperature of $\sim 50$\,K
and a warm component with a temperature of 120\,K peaking
shortward of 60\,$\mu$m \cite{khhs97}. It has been suggested that the
warm component (L$_{\rm warm} = 10^{11} {\rm L}_{\sun}$) is heated by
an active nucleus, while the cold component  (L$_{\rm cold} = 10^{12}
{\rm L}_{\sun}$) is starburst powered \cite{rre93}. 
\subsection{Probability of a chance superposition}
The {\it a posteriori} probability of finding a 60\,$\mu$m source
 within a PHT 41$''$ diameter beam can be estimated by extrapolating recently
 determined ISO survey source counts at 15\,$\mu$m, 90\,$\mu$m and
 170\,$\mu$m. At 15\,$\mu$m the integral source counts at
 50\,mJy from the ELAIS survey are $\sim10^4$ per steradian 
while at 90\,$\mu$m there are 10$^5$ to 10$^6$ sources per steradian to
 the same flux limit \cite{roe+99}. 
In a deep survey of the Lockman Hole,  source counts at
 95\,$\mu$m and 170\,$\mu$m at 150\,mJy extrapolate into the same range as
 ELAIS at
 50\,mJy \cite{ksm+99}. The consistency of the source counts from 90
 to 175\,$\mu$m and the increasing source counts from 15\,$\mu$m to
 175\,$\mu$m indicate that an estimate of $\sim10^5$ sources per steradian at
 60\,$\mu$m is reasonable. This implies a probability of $\sim5\times
 10^{-3}$ of finding a 60\,$\mu$m source down to a detection limit of
 50\,mJy by chance in a PHT beam of 41$''$ diameter. 
 
Alternatively, modelled source counts at 60\,$\mu$m can be used to
derive the probability that the 60\,$\mu$m source is a chance
superposition with the GRB host \cite{prr96}. Taking into account
source counts due to normal, starburst, Seyfert and HyLIGs, 
the probability of finding a 60\,$\mu$m source with
the observed flux within a PHT beam is also $\sim 5\times 10^{-3}$. 
These estimates are sufficiently small to hypothesise a physical 
connection between the 60\,$\mu$m source and the host galaxy of the GRB.
\subsection{The star formation rate of the 60\,$\mu$m source}

It has long been recognised that some starbursts are obscured by dust 
 \cite{ken98} and recent results from ISO have shown that such
 obscuration is widespread and important for the history of star
 formation in the universe \cite{eac+99}. Hence 
far-infrared luminosities provide a more reliable estimate of star
formation rates in galaxies. 
If the source is a ULIG, then most of its massive star
formation occurs in dense molecular clouds and what is
observed at visible frequencies represents emission from stars forming
near the edges of clouds which can escape directly without being
reprocessed by the dust \cite{rrmo+97}.
 The SFR can 
be deduced from the far-infrared luminosity using the following
expressions for M$_*$, the 
rate of star formation per year in units of M$_{\sun}$:
\begin{math}
        {\rm M}_* = 2.6\phi/\epsilon \times 
\frac{{\rm L}_{60}}{{\rm L}_{\sun}} \times 10^{-10} \nonumber
  ~{\rm and}~ {\rm M}_* = 9.3\phi/\epsilon \times \frac{{\rm L}_{15}}{{\rm
L}_{\sun}} \times 10^{-10} \nonumber
\end{math}
~where $\phi$ incorporates a correction factor from a Salpeter initial
mass function (IMF) to the true IMF and a 
correction if the starburst is only forming massive stars \cite{rrmo+97}. 
$\epsilon$ is the fraction 
($\sim 1$) of optical and ultraviolet energy emitted in a starburst
(lasting 1\,Gyr) 
which is absorbed by dust and re-emitted in the far infrared. 
${\rm L}_{60}$ and ${\rm L}_{15}$ are the restframe 60\,$\mu$m and
15\,$\mu$m luminosities ($\nu {\rm L}_{\nu}$ expressed in units of 
${\rm L}_{\sun}$).
These two measures yield values for M$_*$ of between 190$\phi$ and
220$\phi$\,M$_{\sun}$/year. 
An alternative estimate which uses the infrared (8--1000\,$\mu$m)
luminosity and which assumes starbursts lasting $< 10^8$
years, yields a SFR of $\sim 500\,{\rm M}_{\sun}$/year \cite{ken98}.

The identification of a ULIG as the possible parent 
of the  GRB host galaxy therefore favours models of GRB formation involving 
compact stellar remnant progenitors, such as
supernovae \cite{bbh96,woo93} or hypernovae \cite{pac98}, which place 
GRBs in or near star-forming regions. The possible detection of an
iron line in the X-ray spectrum of GRB\,970508
 provides independent support for this scenario \cite{pcf+99}. 
However, the HST data show that the optical transient is well-centred
on the host galaxy, which is quite blue  \cite{fpg+99}. 
Since ULIGs tend to be disturbed, dusty systems, with a significant fraction
undergoing mergers, the host galaxy would be expected to be
 redder in colour and its 
morphology more disturbed than is observed \cite{csms96}. 
Another possibility is that the GRB host 
has been formed by the ULIG, as the product of a merger. 
Small 
galaxies formed along the tidal tails in mergers are likely to detach to
become dwarf systems with characteristics (e.g. colour, absolute
magnitude) similar to the GRB host \cite{sm96,hcz96}. We can speculate
in that case that the offset observed in the May and November 60\,$\mu$m maps 
between the 60\,$\mu$m peak and the OT is genuine, arising from the
observation 
that the most luminous tidal dwarfs are those at the largest projected
distances from the parent nucleus \cite{hcz96}.
The galaxy G1 may also have been formed in the merger but a direct 
redshift determination would be required to support this hypothesis. 
\section{Conclusions}
The ISO observations of GRB\,970508 place unique limits on the level
of far-infrared emission in the weeks and months following the burst
event. Non-transient emission observed at 60\,$\mu$m indicates the
presence of a ULIG which may be the parent of the host galaxy of the GRB. 
This result may have important
implications for GRB progenitor models, favouring those which place
GRBs in or near star-forming regions. 
\begin{acknowledgements}
The ISOPHOT data presented in this paper was reduced using PIA, which
is a joint development by the  ESA Astrophysics Division and the ISOPHOT
consortium.
The ISOCAM data presented in this paper was analysed using CIA, a
joint development by the ESA Astrophysics Division and the ISOCAM
Consortium. The ISOCAM Consortium is led by the ISOCAM PI,
C. Cesarsky, Direction des Sciences de la Matiere, C.E.A., France.
\end{acknowledgements}

\bibliography{mnemonic,iso}
\end{document}